# A Door into Another World[†]

*Abstract*.  Is it possible to design programs which each user can change according to his preferences?  Not an illusion of such a thing that adaptive interface provides but really an interface ruled by users?  What is the main problem of such design and what is the solution to this problem?  This short article gives a glimpse into the theory discussed in the book "The World of Movable Objects".

## Any article has to start with some history

The history of programming is not too long so a lot of readers can easily remember what was before the era of personal computers; the remaining part of readers has no idea about those Ancient Times and what has come from that dark period into our days without any changes at all.  When the term *personal computer* didn't even exist, the overwhelming majority of programs were scientific or engineering; a lot of them were developed for the projects initiated by the Department of Defense, but the tasks themselves and the computer programs to solve those tasks were of the mentioned types.  The researchers who worked on those projects were mostly proficient in math; they easily learnt FORTRAN and turned their knowledge into algorithms and the code of working programs.  At those days it was common that the same person was simultaneously a researcher, an author of the code, and the analyser of results.  The output of the programs was mostly in numbers and could be easily changed by this person.  The algorithms were huge; the programs had long listings and were often written in *spaghetti code*.[*]  In a while you could see more specialization in programming area with different people being researchers and program developers, but everything else was the same.  Researchers and program developers often worked in close collaboration and the problems of the output were solved immediately or after a short discussion on a personal level.  The main thing was that at that time the developer of a program had total control over the calculations and over the visualization.  It was going this way for so long that became an axiom.

Things began to change rapidly with the beginning of PC era and in our days users of some programs can outnumber the developers by thousands or millions; the solving of interface problems through personal communication became inappropriate.  It's impossible to imagine that thousands or millions of users would be absolutely satisfied with the designer's ideas of interface even if it is excellent.  The solution to this problem was obvious: give users an instrument to adapt the interface to their demands.  For the last 30 years the design of interfaces for computer programs was influenced mostly by the ideas of *adaptive interface*.  The computer revolution tremendously increased the number of developers, so the stream of different ideas for adaptive interface was enormous.  Not all of these ideas became popular but some of them are; for example, dynamic layout promoted now as the mainstream for today and for the nearest future.

The adaptive interface gained its popularity many years ago; the results of the research work in this domain were demonstrated in hundreds (more likely - thousands) of papers; it became the dominant religion in the area of interface design.  Any religion has its holy books which anyone has to be familiar with; for the area of interface design the [1] is one of them.  Articles of this book include references to hundreds of previous works; I certainly don't want to copy those lists here, but they can give to any reader a good estimation of how much work was done previously in the attempts to design some kind of interface that would satisfy all users.  All that work resulted in an outstanding progress of interfaces, but, as often happens to some theory that becomes dominant, at one or another moment it turns from being progressive and helpful into a restricting one, because it rejects everything that is not declared in its holy books.  Then comes the time of Reformation.

Each time and area of knowledge requires its own way for the demonstration of new ideas.  I have written a book about the new ideas in the design of applications [2]; this book is accompanied by a huge demonstration program with more than 100 examples.  Some of these examples are solid programs in their own right and are used throughout the scientific work in our department of Mathematical Modelling.  It would be impossible to understand and estimate the novelty of new programs without trying them.  The older readers would remember that it was impossible to understand the novelty of Windows system without seeing it in work; the step from the currently used programs to the new *user-driven applications* is even bigger (from the users' point of view) then that old step from DOS to Windows.  So try to estimate the further text in parallel with running those examples.

---

[†] This article was rejected by Communications of the ACM in November 2011.
[*] Don't mix this term with *spaghetti westerns* which also flourished at that time.  Dijkstra's letter to the Editor of Communications of the ACM, published in March 1968, marks the beginning of structured programming.  I doubt that anyone in his clear mind would like to return to the type of programs that were popular before that, while a lot of people (and I am among them) would like to find a spare time, sit down, and admire once more *The Good, the Bad and the Ugly* (Sergio Leone, 1966).

## What is wrong with the adaptive interface?

The structure of nearly any program can be divided into three main parts: input data preparation, calculations on this data, and the demonstration of results. There are some programs which look like lacking the middle part of this trilogy (take some data from a database and simply show it), but I'll demonstrate further on that even such applications fit very well with the ideas of user-driven applications. The next table shows the distribution of user / developer responsibilities for three stages of the currently used programs and for applications of the new type.

|  | Input data | Calculations | Output (results viewing) |
|---|---|---|---|
|  | R e s p o n s i b i l i t y | | |
| Currently used programs | User | Developer | Developer |
| User-driven applications | User | Developer | **User** |

**Table 1**. Responsibility for the stages of the currently used programs and user-driven applications

The boundaries between the user/developer responsibilities were very strict decades ago but became fuzzy in our days. When some form of adaptive interface is used in a program, then it looks like users have some influence over the output or even control it. In reality adaptive interface gives only an illusion of users' control; it never gives user an opportunity to do whatever he personally wants, but only allows to select among choices which were previously considered by developers as good enough for one or another situation. This is a fundamental law and the main flaw of adaptive interface. If it happens so that some user shares the developer's ideas about the best looking interface, then this user is really lucky and can get exactly what he wants. But it is a very rare situation. More often than not you either hate the interface of a program or strongly dislike it but in any case you have to work with it because you are more interested in the results that you can obtain from this program. Adaptive interface is always called a friendly interface regardless of whether you like the introduced example of it or strongly dislike. It looks like a mockery to me.

The good developers are well aware of those interface problems and try to solve them. Unfortunately, they continue to look for a new solution in the standard way: if users are not satisfied with some part of a big program, then this part is redesigned in such a way as to give users more flexibility in this segment. Simply a well known or a new and more sophisticated version of adaptive interface is used. It's exactly like trying to develop a real perpetual motion machine by using new materials which were not available years or decades ago. One more effort and we'll get it!

The most popular form of adaptive interface used in our days is the dynamic layout. Try to say a word against the dynamic layout and you immediately receive a snub from the authorities on user interfaces. *"Dynamic layout is actually motivated by usability concerns, not just by developer convenience. Dynamic layout allows an end-user to resize a container (such as a window or dialog box) and have the application automatically resize and rearrange the elements inside the container. Without dynamic layout, the end user would have to manually, one by one, resize and reposition the elements inside. So dynamic layout does confer usability benefits by making the user interface more efficient: one resize action by the user results in many automatic resizes and repositions of dependent objects."* (Personal letter from Robert Miller, October 2011). This description emphasizes the pluses of dynamic layout; it wasn't supposed to underline its minuses but it did. From my point of view the minuses of using dynamic layout significantly outweigh those pluses. With the implemented dynamic layout, users get the easiness of resizing but in exchange they have to give away all the possibilities of changing the inner view of an application in the way they want. They are left no chances: the application will always look according to the developer's ideas. It's an extremely high price: you give away the freedom and the ability to make any decisions on your own for a daily ration. What is more, nobody ever asks users if they want such an exchange or not. The dynamic layout is easy to implement and developers implement it whenever they want. A classical idea of modern development: "You would have to like whatever we are giving you".

There was a period in the history of programming when the code for calculations was intermixed with the code for visualization and this was considered normal; later two parts of code were separated and the benefits of such decision were high. Up till now both parts continue to be under developers' jurisdiction; it's the time to separate also control over them. The idea of taking the control over visualization (over output results) from developers and giving it to users looks absolutely heretical to nearly all developers. The prevailing view among developers about users of their programs is short and never publicly expressed "Users are idiots". I heard it not once in private professional discussions and every time I laughed because I made simple extrapolation and then the developers in Microsoft must express the same opinion about all those using Visual Studio in their every day work.

As shown in the table, the difference between the currently used programs and the new applications is only in the last column. Is this difference really so important? In nearly any application the stages of calculation and results viewing are not organized one after another but are going simultaneously. By watching the outcome of results user often makes a decision to interrupt the process, to change the input data, and to start the calculations again. So in this way user keeps the full control of application. If the control over the output is so important, then in what way it can be produced? The importance of switching the control over the results viewing from developer to users can be and must be demonstrated on the real programs; we'll come to it a bit later and now let's look at the conditions to provide such a switch.

## What is the basis of user-driven applications?

*"In science, finding the right formulation of a problem is often the key to solving it..."* [3].

For several decades I work on design of very sophisticated programs in different areas of science and engineering. Though the areas are far away from each other (speech analysis, telecommunication, analysis of big electrical networks, thermodynamics…), I found out that the development of those applications has a lot of similarities and throughout the years I began to feel that there was one general problem. Design of the most popular programs in each particular area hardly changed at all throughout the last 12 – 15 years and I saw that the same thing happened not in one but in all those areas. I began to think about the cause of this general stagnation. It turned out that the problem was in the total developers' control over the applications. The programmers are professionals in their area of development; they can be very high professionals, but they are not as good in understanding the problems of each particular area as the researchers and engineers. Yet, those researchers have to work with the programs that are developed according to the level of developers' understanding. It means that the better specialists are limited in their work by the lesser specialists and there is no way around this problem if the designing of the programs continues in the standard way. There must be some way to change the overall design of applications in order to avoid this domination of the lesser specialists over the research work, but where is this way?

The solution to this general problem is in turning the currently designed applications, which are in reality not more than sophisticated calculators for predetermined situations, into something different, into real instruments. The difference may be not obvious at the first moment but it is fundamental. Instruments have no detailed instructions on their use. For example, a Calculator is an instrument; it has no detailed list of numbers that can be multiplied or added; Calculator can be used for any real numbers. The developer gives an instrument – Calculator, but only user decides about the exact numbers for which it is used at one moment or another. The same thing has to be done with every application: programmers develop an instrument and users decide about its use.

The fundamental interface problem is based on the assumption that an application must be run by developer. It was so in the old times; there was a chance to change the trend and make the right (different) decision at the moment when the multi windows operating systems were born, but another decision was made. At the level of operating system users have the full control over the elements. Users can move and resize windows; they can open the new windows, close the unneeded, and place everything in whatever way they want. Users can also move the icons around the desktop. There are no other elements at that level; users can do whatever they want with all the existing objects, so they have total control at this level. But start any application and try to move any object inside. Is it possible? With an exception of few programs (like *Paint* and similar) nothing can be moved inside the applications. At the inner level we have not only rectangular objects but elements of the arbitrary shape; to organize users' control over all the inner elements you need some algorithm of turning any element into movable. Such algorithm was not invented and this path was closed. The door was shut and there was no key.

Instead of the full users' control over the applications the decorations were developed in the form of adaptive interface. Combining efforts of many researchers and developers throughout the next 25 years produced the amazing decorations. The whole programming world lives inside these decorations for the last 25 years and the majority of developers were even born inside this environment, never saw anything else, and were taught on the books declaring this environment as the only real thing. Nobody declared that programs can be developed in a different way and the currently used idea of developers' control became an axiom. It always was, it works this way in all currently used programs (except mine!), and it always will be! Are you sure? It's not a nature law like a law of universal gravitation. The developers' control over interface is a historical fact and nothing more. It is based only on the currently used programming technique, but what if you change this basis?

I don't see or expect any objections to the switch of control over programs from developers to users, because from the users' point of view it would be a huge benefit. <u>The total control over an application must be given to users</u>. Now it looks obvious to me and I can only ask myself, why it wasn't obvious to me years ago and why no one else came out with the same solution before. It's the normal situation in science: you think a lot about some problem and when the solution is found, you can only shrug the shoulders, because it's the only good solution and had to be obvious from the very first moment.

This total control over applications means that any screen object regardless of whether it is a primitive element or a complicated object consisting of many parts must be turned into movable / resizable and the full control over those movability and resizability must be given to users. It is not the movability of objects by the designer – this would be a simple animation which has nothing to do with the discussed problem. We need some algorithm which would allow any user to press a screen object and either move it or resize it whenever it is needed.

Were there previous attempts to provide the movability of objects by users inside the working programs? Every professor teaching courses on interface design like to say that this is a well known thing and that it was implemented years ago, for example, in the Morphic system. In reality this statement is absolutely wrong. You can take a good description of Morphic [4] and find out that that iconic system gives only three possibilities for lining the elements (page 22; "A justification parameter controls placement in the secondary dimension; for example, the tops, bottoms, or centers of a row's submorphs can be aligned with the top, bottom, or center of the row."). There is no arbitrary movability and placing of elements by users; there are only three choices of lining provided by the developers of that system. This is a classical case of adaptive interface with some variants considered to be good by developers; only these variants are coded and allowed. Somebody wrongly mentioned years ago that Morphic implemented the movement of elements (authors of the system never did it; it was somebody else!) and this wrong statement, after numerous repeats, became a well-known fact. A classical falsification of history!

Were there other attempts? Certainly, but all of them were limited to one or another special cases. From time to time you need the movable objects in your application and good programmers organize movability for each particular case. For each class of movable objects the special algorithm was used; those algorithms were complicated and never used for anything else. I have done it 15 and 20 years ago in some of my programs and those old algorithms could not be used for anything else except those special cases of specific objects.

The new type of programs – user-driven applications – can be designed only on the total movability of all the screen elements and the algorithm to provide this movability must have two very important features. The first requirement comes from developers who are going to use this algorithm; the second one – from the users of such applications.

- An algorithm to turn any object into movable / resizable must be easy to implement for all objects without exceptions.

In the book [2] I described an algorithm of turning any screen object into movable / resizable. The algorithm is based on covering an object by the invisible areas; each area is responsible for moving, resizing, or reconfiguring an object; the whole process is done with a mouse. The main requirement was the easiness of applying the same algorithm to any object of an arbitrary shape and origin; up till now I never met an object which would cause a problem with making it movable. Even in the few examples which are shown further on in this article you can see very different objects, but the book demonstrates much more. There are geometrical figures of different shapes; there are elements with holes; there are different types of groups; there are complicated objects with the parts involved in individual, synchronous, and related movements. The variety of screen objects is infinitive and I wasn't planning to show the solution for each particular case. It's like the mathematical analysis: you have to understand the ideas of differentiation and integrals; you have to know several often used techniques, and on this basis you can start to solve any of your problems. Everything else depends on your brains, your skills, and the spent time.

- Moving and resizing must be easy and similar in use for all objects.

From the users' point of view these similarity and easiness are provided by two general rules: any graphical object can be moved by any inner point and resized by borders. The rules for controls are slightly different as their inner area is used for standard commands and users do not expect anything different there. Thus, controls are moved and resized by different parts of their border.

When you have an algorithm for turning any screen object into movable / resizable and start designing programs on the basis of such elements, you quickly come to the understanding that these new programs – user-driven applications – have very strict rules which is impossible to ignore.

- All the elements are movable.

- All the parameters of visibility must be easily controlled by users.

- The users' commands on moving / resizing of objects or on changing the visibility parameters must be implemented exactly as they are; no additions or expanded interpretation by developer are allowed.

- All the parameters must be saved and restored.

- The above mentioned rules must be implemented at all the levels beginning from the main form and down to the farthest corners.

These rules do not depend on the particular algorithm; you can use my algorithm or design something of your own; in any case you come to the same rules. They are like nature laws; you try to ignore them or break them and the logic of application will demonstrate you that you are wrong and it's much better to design according to these laws. Now let's turn to some examples and see how these rules work in real applications.

**User-driven applications**

The first example is the *Calculator* program which, if I am not mistaken, appeared in the very first version of Windows system and never changed since that time; left picture from **figure 1** demonstrates the view of this program. I am sure that this view is well known to any reader of this article and nearly every reader will ask a simple question: "What is wrong with this application? It is used by millions of people throughout the years and there are no complaints." I absolutely agree with the first part (it is used by millions), but I am not so sure about the absence of complaints. At least, my work on *Calculator* was triggered by such complaint from one of the colleagues. With the increasing distance between the current day and the day of birth, people prefer to use the bigger font for the screen objects, but all the attempts to change the font in the classical *Calculator* showed an amazing result: there is no way to do it. To this I can add my own problem. I use this *Calculator* not more than three or four times a year; I prefer to make all the calculations either in head or with a pen on a sheet of paper – old people prefer to keep to the old habits. Each time I use the *Calculator*, it strikes my mind that the needed numbers are definitely not at the places where I would like them to be. I have to find the needed digits; it takes only an instant, but it is annoying. When I designed for my colleague another *Calculator* which allowed to change the font, I certainly designed this *Calculator* as a normal user-driven application, so all the elements in view are movable, resizable, and tunable. Two right pictures at **figure 1** demonstrate only two possible views from the infinitive number of variants.

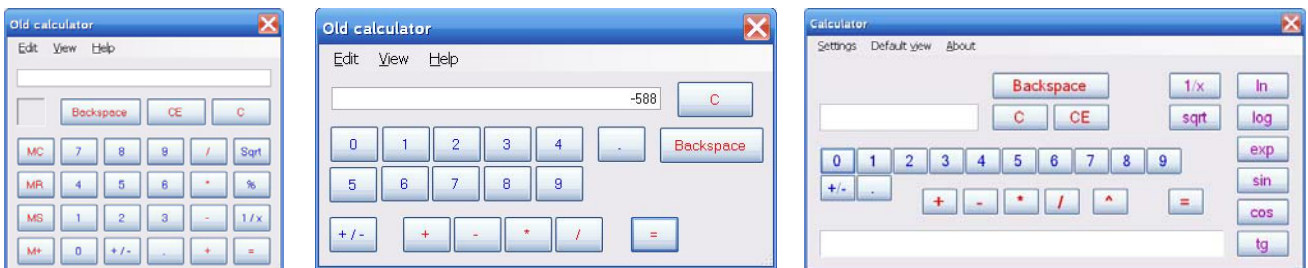

**Figure 1**. Three views of the *Calculator* program: the classical one, slightly changed, and redesigned.

This is definitely not the case of adaptive interface with the fixed number of prepared variants. This is also not the dynamic layout which would never give you such flexibility. Each user can change the view of this *Calculator* in any way he wants, and at different moments the preferable view can be different even for the same person. *Calculator* works correctly regardless of the places, sizes, and views of all the inner elements. Developer has to provide the correct reaction on the click of any button and the correctness of calculations, but the view of the program is absolutely under users' control and developer must be banned from interfering with these users' actions. Developer must think about the best possible design from his point of view and provide an easy way (one - two clicks at maximum) to reinstall this default view at any moment, but it is one more option for users and nothing else. The view of the application is under the full user's control.

On my proposal for the UIST-2009 conference one of the reviewers (unfortunately, anonymous) wrote such a thing: "*Is there more than anecdotal evidence to suggest that end users really want to be configuring the fine points of a user interface layout, rather than accomplishing the task that the application is intended to support? (The HCI literature would seem to provide strong evidence to the contrary.)*"

First, even if something is written in your holy book and considered to be an axiom at the moment, it doesn't mean that it is really true. The famous doctrine declared for centuries that the Earth was the center of the Universe; it turned out to be not correct. Second, the movability of elements doesn't conflict with the correctness of *Calculator*. There is no choice between the correctness and movability. There is no selection between making the calculations or moving the screen elements. Users need this program for calculations and the program works according to its main purpose, but with an addition of elements' movability users get a chance to organize the program in the best possible view for each of them.

I am sure that you can find approximately the same set of books with only minor variations in the offices of the professors teaching the interface design in numerous universities. The books are the same but the placement of those books is very individual and depends on the room size, number and size of the bookshelves, light, and the habits of each owner. I am sure that you will never find exactly the same order and placement of the books even if you spend a lot of time on looking into those offices. Try to tell those professors that they have to accomplish the task of education and you, as the best specialist on room design, will place all their books in the way which would be the most efficient for their work. I doubt that anyone would ever publish even the most polite words that you would hear in response. Then why the same professors continue to declare that users don't need a chance to rearrange the view of a program but need only to think about accomplishing the task?

If we agree that movability of elements inside the programs can be useful (I hope that now you don't reject this idea as heresy), then there arise those questions of simplicity to organize such a thing and to use it.

<u>On the designer's side</u> it is organized in such a way. All the elements in this *Calculator* are controls; at the design stage all of them get the range of their resizing. Then an additional object (`mover`) of the `Mover` class is declared; this object supervises the whole moving / resizing process.

```
Mover mover;
```

All the elements to become movable – all the controls in the form – are registered with this mover.

```
foreach (Control control in Controls)
{
    mover .Add (control);
}
```

I decided (it looks the best way for me) that the whole moving / resizing must be organized only with a mouse (press – move – release), so only three mouse events are used. This is the only needed addition to the code of *Calculator*. You can estimate for yourself if the following piece of code is too complicated to organize an unlimited flexibility of *Calculator*.

```
private void OnMouseDown (object sender, MouseEventArgs e)
{
    mover .Catch (e .Location, e .Button);
}
private void OnMouseUp (object sender, MouseEventArgs e)
{
    mover .Release ();
}
private void OnMouseMove (object sender, MouseEventArgs e)
{
    if (mover .Move (e .Location))
    {
        Update ();
    }
}
```

As I have already mentioned, <u>users are doing</u> all moving / resizing simply by a mouse. Each control can be moved and resized individually, but there are so many elements that this would be not enough. There are three predefined groups of elements – buttons for numbers, operations, and functions are shown at the right picture at **figure 1** in different colors. Each of these groups can be looked at as a single object; the position of a group and the visibility parameters of their elements can be changed by a single command. In addition you can draw with a mouse a temporary rectangular frame around any group of elements and deal with this temporary group in the same simple way; there are several lines of code to organize such a group, but this code is as simple as shown above.

*Calculator* is the first example which I demonstrate in this article. I want to mention how the general rules of user-driven applications are applied to this program and to underline that they are applied in exactly the same way to all other programs of the new type. Also those rules, mentioned two pages ahead, are linked with each other and if you implement the movability of elements into your program, then you automatically come to other rules regardless of whether they are declared or not.

Movability means that each element can be placed at an arbitrary place. If user organized the preferable view of a program, then this view must be saved on closing an application and restored the next time it starts. The same thing happens with all the visibility parameters. There are not too many of them for each element: position, size, color, and font. The time when we were thinking about each bit and byte in memory or on the hard drive are long gone; the parameters must be saved (and restored) for each element individually. As a developer, I may think that all buttons with numbers must be of the same size and use the same font and color, but some users may think differently, and I have no right to interfere with their decisions. If they want to enlarge some buttons or to set the personal font or color for some of them, I have to provide an easy way to do it and even think a bit further on. With the decades of interface design, I can think out a lot of possibilities, but I am going to use this experience only as a help, as another possibility but not as a restriction. I don't impose my preferences over the users' possible actions but simply give them as possible variants. For example, I provide an easy way to set some standard views for groups of numbers, operations, and functions; I also add into the context menus the commands to declare the view of the touched element as a sample and to spread this sample view on all other elements of the same group or even wider. The number of the available commands and the simplicity of their use depend on the developer's experience, but the most important thing is the understanding that none of these commands is the overruling of developer's preferences over the user's decisions.

The next example is the program which deals with personal data information. It's a widely used case that a lot of personal data is stored somewhere in the database and people need an application to look through this data and to change it from time to time. The designers of such programs think about the best possible way to show the information and about the possible views when somebody needs not the whole set of data but only a part of it. Designers put huge efforts into organizing those several good views and are really proud of their results. On top of all these efforts there is a survey of users' satisfaction and the loud announcement that "a new version showed the increase of users' satisfaction from 83 to 84.5 per cent". What these numbers really show is the per cent of those who agree to work with whatever they are given because the users are interested in dealing with information from the database and are ready to work with any interface which shows the information without crashing every 30 minutes. Users adjust to whatever they are given because they have no other choice; they have to work with the important information regardless of the way in which it is shown.

Just try to remember how many times you were mad with the new view of popular applications. Each new version is declared as the revolutionary step in the history of mankind, while you desperately try to find how to do those simple things which you did without problems in the old version. If there are no kids near by, you may pronounce about the designers some words which you would prefer the kids not to hear. I think that the designers are very good, maybe some of the best, and still people around the world are mad with their results. It's not the level of design; it's a perfect demonstration of a huge variety in opinions on how the programs must be organized. There are the best achievements in adaptive interface, there is that famous dynamic layout, and there are millions of people around the world who are simply mad with the proposed design. In 1775 the Royal Academy of Sciences in Paris issued the statement that the Academy "will no longer accept or deal with proposals concerning perpetual motion". In which year to come the developers will understand that no version of adaptive interface can produce a decision which is excellent for each particular user? Adaptive interface is a popular highway to a dead end.

Switch the main goal and the efforts. Instead of imposing on everyone the design that you think must be good for everyone, give such view as a default one (no questions, you have to develop it on the best possible level) but also give an easy to use instrument for any user to change the view in whatever way he wants. That's the main idea of user-driven applications.

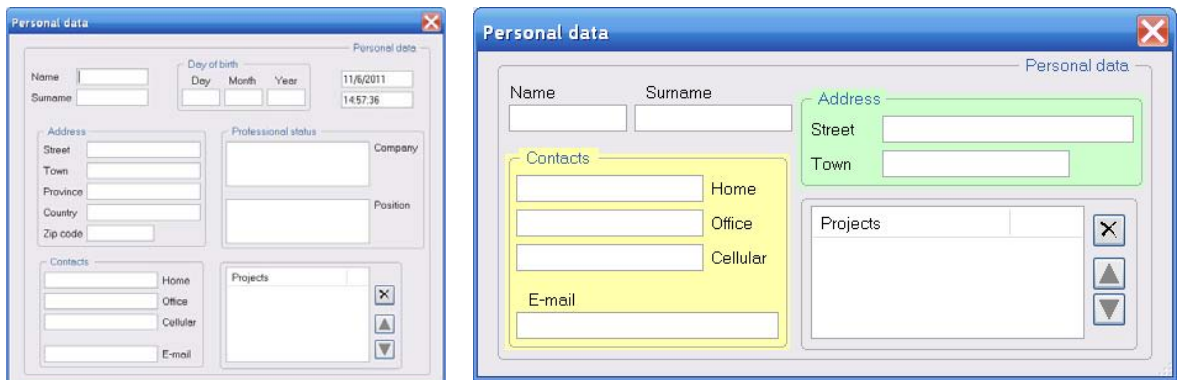

**Fig.2**  The default and the changed views of *PersonalData*

**Figure 2** shows the default and the customized view of the *PersonalData* application. All the data is shown inside one big group which contains five smaller groups and controls with and without comments; the inner groups also contain different types of controls with and without comments. The default view shows the full set of elements which were included into this application. At any moment user can hide the unneeded groups and elements and concentrate his attention only on whatever is really needed; hiding of the currently unneeded objects and restoration of the elements or groups later is done by a single command in an appropriate context menu. I demonstrate on these pictures only a glimpse of the opportunities for changing the visibility parameters; colors and fonts can be used to highlight the needed information. By the way, movability of the elements allows to solve a problem which would in other systems require a significant code addition and even this would not guarantee the correct result. Not all countries in the world use the address information in the same order; there are countries in which the address starts with the ZIP code, then comes the name of the country, and so on; in this application you can move the elements inside the groups in any preferable way, so you can change the order of elements inside the *Address* group in a second.

The groups, which are widely used in this application, can be of special interest by themselves, because they have a lot of interesting features, but they also bring attention to the fact that by adding the movability to some well-known element you may not only add something new to this particular element but you can open an area of absolutely new ideas. The main purpose of using groups is to demonstrate that several related elements are used for one task or for changing the related parameters. For years we have only two options to organize the groups: there are panels and there are standard groups with a title and a frame around inner elements. The frame or the panel's border indicates a set of elements included into a group. When the dynamic layout is applied to a panel, then the resizing of this panel changes the sizes of the inner elements. The purpose of the border is only to indicate the belonging of elements to the collection, but when the dynamic layout is applied, then the auxiliary element – border – begins to rule everything. It's the logic turned upside down. The main elements are the inner elements of any group; they show the information, they start any action, and they have to be the ruling elements of a group.

This idea is implemented in the groups from **figure 2**; such groups of the `ElasticGroup` class became the main constructing elements in the majority of my applications. According to the rules of user-driven applications, all elements in the groups are movable and resizable; user can do whatever he needs with those elements and the frame around the group will always adjust to all the inner changes. It's an elastic frame around the collection, but it also indicates the area which can be moved by any inner point.

Once more about the easiness of changes. There can be a huge number of users for such application and each user would organize the view which is the best personally for him and according to the part of information that he has to deal with; let's say that this is the right view from **figure 2**. Then comes the time of the Christmas cards when you need the full address for each person; for this you simply unveil the currently hidden elements in the *Address* group (one command in menu) and the database will show you all the needed information. After the holidays you rearrange the view again. Your do not do it all the time, but you can easily do it at any need.

The work with the new applications can demand from designer something that he wasn't even thinking about. For a long time I was mostly thinking about an easy way of turning any element into movable / resizable. When such applications began to work, their users immediately demanded an easy way (one-two clicks) to change the status of any element and group from movable to fixed and back. When you work with a densely populated group, like *Address* group on the left picture, chances are high that on trying to move the whole group you accidentally move some inner element; fixing of elements solves the problem.

I want to demonstrate that absolutely different applications can benefit from being turned into user-driven. The first example – *Calculator* – contains only controls; the second – *Personal data* – contains a combination of controls and graphical elements; the last example – *Functions analyser* – is based mostly on the graphical objects with only a small addition of controls. There are several different forms in this application and **figure 3** demonstrates only one possible view of the main form.

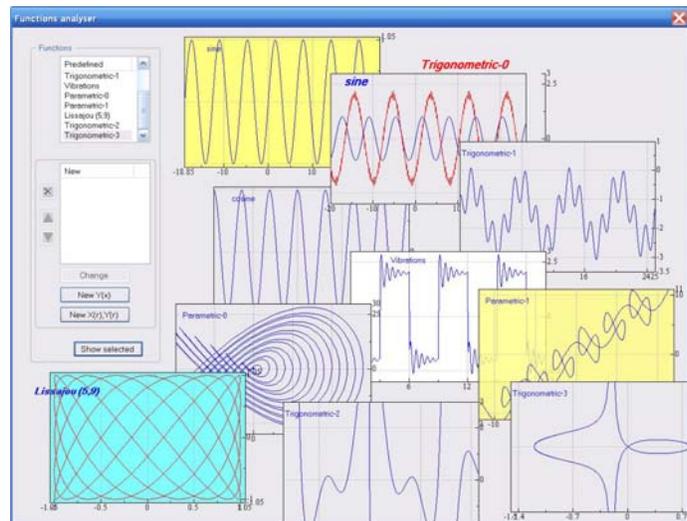

Suppose that you teach a course on mathematical functions. You are going to talk about trigonometric functions, exponents, logarithms, and some others, but for any of those functions you would like to demonstrate how coefficients and additional parameters can change the view. At any moment your students can ask you to compare those functions with something new and not predefined. If you would be teaching a course, you would organize the view in the way you need; **figure 3** only reminds me about the set of predefined functions.

The preliminary discussion of a future application between the developers and would be users (scientists, engineers) was never an easy process and was often accompanied by quarrels. All the books on software engineering declare the first rule of design: never start any development of a big program until the specification is thoroughly discussed and every detail is protocoled. If you start without it there will be an infinitive stream of changes which will result in patches over patches and never ending development. Unfortunately, this rule sounds perfectly but can't be used in real life. By the core nature of research work nobody can predict the future outcome in a month or two from today and the requirements to the used applications are going to change in parallel with the research.

**Fig. 3** Functions analyser

For years I was developing applications according to the users' requirements; in the standard version of the same program I would have to discuss the number of possible plotting areas, their positions in each case, and would have to think about not a trivial technique to change this number on the fly. For each number of plotting areas I would have to predetermine the best possible configuration and then I would have to develop some mechanism for users to change a view. Not in any way they want but inside the limitations that I would have to impose on them. A classical type of adaptive interface! I was developing such systems 20 years ago. I wouldn't call such development a nightmare; it was really an exciting thing to design those systems. Users even liked them (maybe because they had nothing better), but in comparison with the scientific applications that I give to users today it was a masterpiece of the Stone Age.

Now you teach the course and you decide WHAT, WHEN, and HOW to show on the screen or maybe even on several screens. Fot the lecture on trigonometrical functions you organize one set of plotting areas with their scales and comments; for the next hour on exponents you have absolutely different view. My work, as a developer, to provide you with an Interpreter that allows to type in any function in standard mathematical notation and to provide an easy to use instrument of setting any parameters of visibility. There is also the saving and restoring, but this is a standard feature of any user-driven application. You are the user – you are driving; my work is absolutely hidden from view – all the calculations are deep inside and developer is now responsible only for calculations. The time have changed!

This is an easy to understand application as everyone is familiar with the commonly used functions, but the main rules for design of this application are the same as for other very specific scientific applications. I wrote in another article [5] that researchers work now with the applications of the new type exactly as they were working with the sheets of paper years ago. They can organize any number of plots to see some results and delete them at any moment if they are not needed any more. The plots can be placed side by side for better comparison or in any other way; any plot from underneath can be called on top at any moment. Plus there are all the possibilities for zooming, printing, changing the auxiliary information, changing the visibility parameters of each and all elements, saving, restoring… Scientists are never satisfied with whatever they are given; they always have some ideas on improvements. Some of these ideas would be considered absolutely crazy before the era of user-driven applications; some of these ideas sound crazy for me even now, but this makes my work only more exciting.

## Conclusion

I think that the time have come to explain the title of this article. Years ago Clifford Simak wrote "*The Big Front Yard*" for which he received the 1959 Hugo Award. The story is about the house in which an ordinary door opens the way into another amazing world. On one side of the door there is our normal world with all the familiar features; on another side there is the *terra incognita* which has to be explored.

I have spent my whole life on design of engineering and scientific applications. Years ago I understood that somewhere 12 – 15 years ago the whole area of such applications went into stagnation; I started to look into the cause of it. Thoughts about the real cause of this stagnation brought me to some conclusions and sparked the work on the general algorithm for movability of the screen objects. My long and numerous searches throughout the past results only proved that there were no previous successful attempts to build a system or a big enough application on the basis of movable objects and with a full control passed to users. As one of my colleagues liked to describe similar tasks years ago: "It is impossible because it is never possible."

Yet, it turned out to be possible and produced the outstanding results. The algorithm works now in applications for many different areas; this proves only the universality of this algorithm. But the algorithm itself is not the main result of my work; it only allows to open a door into another world. The idea and implementation of user-driven applications is much more important. You can use my algorithm or design something of your own. I am sure that other algorithms for moving all the screen objects will be produced, but any algorithm will be a foundation for the same type of programs which I call *user-driven*. The main idea is in passing the full control over applications to users and this can be done only when users can do whatever they want with all the screen objects. The door is opened; there is a whole world on the other side of it and even the first results of exploration are extremely exciting.